# Proper projective collineation in non-static spherically symmetric space-times

Ghulam Shabbir and Amjad Ali

Faculty of Engineering Sciences,

GIK Institute of Engineering Sciences and Technology,

Topi Swabi, NWFP, Pakistan

Email: shabbir@giki.edu.pk

**Abstract**

We investigate the proper projective collineation in non-static spherically symmetric space-times using direct integration and algebraic techniques. Studying projective collineation in the above space-times, it is shown that the space-times which admit proper projective collineations turn out to be very special classes of static spherically symmetric space-times.

## 1. INTRODUCTION

The projective symmetry which preserves the geodesic structure of a space-time carries significant information and interest in Einstein's theory of general relativity. It is therefore important to study these symmetries. The aim of this paper is to find the existance of proper projective collineation in non-static spherically symmetric space-times. Different approaches [1-12] were adopted to study projective collineations. In this paper an approach, which basically consists of some algebraic and direct integration techniques, is developed to study the projective collineation for the above space-times. Doubtless, this approach is lengthy but it will definitely tell the existance of a proper projective collineation.

Throughout $M$ represents a four dimensional, connected, Hausdorff space-time manifold with Lorentz metric $g$ of signature (-, +, +, +). The curvature tensor associated with $g_{ab}$, through the Levi-Civita connection, is denoted in component form by $R^a{}_{bcd}$, and the Ricci tensor components are $R_{ab} = R^c{}_{acb}$. The usual covariant, partial and Lie derivatives are denoted by a semicolon, a comma and the symbol $L$, respectively. Round and square brackets denote the usual symmetrization and skew-symmetrization, respectively.

Any vector field $X$ on $M$ can be decomposed as

$$X_{a;b} = \frac{1}{2} h_{ab} + F_{ab} \tag{1}$$

where $h_{ab}(= h_{ba}) = L_X g_{ab}$ and $F_{ab}(= -F_{ba})$ are symmetric and skew symmetric tensors on $M$, respectively. Such a vector field $X$ is called projective if the local diffeomorphisms $\psi_t$ (for appropriate $t$) associated with $X$ map geodesics into geodesics. This is equivalent to the condition that $h_{ab}$ satisfies

$$h_{ab;c} = 2g_{ab}\eta_c + g_{ac}\eta_b + g_{bc}\eta_a \tag{2}$$

for some smooth closed 1-form on $M$ with local components $\eta_a$. Thus $\eta_a$ is locally gradient because the connection is metric and will, where appropriate, be written as $\eta_a = \eta_{,a}$ for some function $\eta$ on some open subset of $M$. If $X$ is a projective collineation and $\eta_{a;b} = 0$ then $X$ is called a special projective collineation on $M$. The statement that $h_{ab}$ is covariantly constant on $M$ is, from (2), equivalent to $\eta_a$ being zero on $M$ and is, in turn equivalent to $X$ being an affine vector field on $M$ (so that the local diffeomorphisms $\psi_t$ preserve not only geodesics but also their affine parameters). If $X$ is projective but not affine then it is called proper projective collineation [2]. The vector field $X$ is said to be proper special projective collineation, if $X$ is not affine and $\eta_{a;b} = 0$. Further if $X$ is affine and $h_{ab} = 2cg_{ab}, c \in R$ then $X$ is homothetic (otherwise proper affine). If $X$ is homothetic and $c \neq 0$ it is proper homothetic while if $c = 0$ it is Killing.

The second order skew symmetric tensor $F_{ab}$ is called a bivector (at p). Regarding $F_{ab}$ as a skew matrix, its rank is therefore an even number 0, 2 or 4. If it is 0 then $F_{ab} = 0$. Suppose if the rank of $F_{ab}$ is 2 then it is called simple bivector otherwise it is called non-simple (for more detalis see [13]). Here, at $p \in M$ one may choose a orthonormal tetrad $(t, r, \theta, \phi)$ satisfying $-t^a t_a = r^a r_a = \theta^a \theta_a = \phi^a \phi_a = 1$ (with all others inner products zero). Since at $p$, the set of bivectors at $p$ is a six-dimensional vector space which can spanned by the six bivectors given by [13]

$$^1F_{ab} = 2t_{[a}r_{b]}, \quad ^2F_{ab} = 2t_{[a}\theta_{b]}, \quad ^3F_{ab} = 2t_{[a}\phi_{b]},$$
$$^4F_{ab} = 2r_{[a}\theta_{b]}, \quad ^5F_{ab} = 2r_{[a}\phi_{b]}, \quad ^6F_{ab} = 2\theta_{[a}\phi_{b]}.$$

In general, however, equation (2) is difficult to handle directly and alternative techniques are needed. One such technique arises from the following result. Let $X$ be a projective collineation on $M$ so that (1) and (2) hold and let $F$ be a real curvature eigenbivector at $p \in M$ with eigenvalue $\lambda \in R$ (so that $R^{ab}{}_{cd}F^{cd} = \lambda F^{ab}$ at $p$) then at $p$ one has [14]

$$P_{ac}F^c{}_b + P_{bc}F^c{}_a = 0 \qquad (P_{ab} = \lambda h_{ab} + 2\eta_{a;b}). \tag{3}$$

Equation (3) gives a relation between $F^a{}_b$ and $P_{ab}$ (a second order symmetric tensor) at $p$ and reflects the close connection between $h_{ab}$, $\eta_{a;b}$ and the algebraic structure of the curvature at $p$. If $F$ is simple then the blade of $F$ (a two dimensional subspace of $T_pM$) consists of eigenvectors of $P$ with same eigenvalue. Similarly, if $F$ is non-simple then it has two well defined orthogonal timelike and spacelike blades at $p$ each of which consists of eigenvectors of $P$ with same eigenvalue but with possibly different eigenvalues for the two blades [15].

## 2. Main Results

Consider a non static spherically symmetric space-time in the usual coordinate system $(t, r, \theta, \phi)$ (labeled by $(x^0, x^1, x^2, x^3)$, respectively) with line element [16]

$$ds^2 = -e^{A(t,r)}dt^2 + e^{B(t,r)}dr^2 + r^2(d\theta^2 + \sin^2\theta \, d\phi^2). \tag{4}$$

The Ricci tensor Segre type of the above space-time is {1,1(11)} or {2(11)} or one of its degeneracies. The above space-time admits three linearly independent Killing vector fields which are

$$\cos\phi\frac{\partial}{\partial\theta}-\cot\theta\sin\phi\frac{\partial}{\partial\varphi}, \quad \sin\phi\frac{\partial}{\partial\theta}+\cot\theta\cos\phi\frac{\partial}{\partial\phi}, \quad \frac{\partial}{\partial\phi}. \tag{5}$$

The situations are well clear when the above space-times admit proper affine vector fields and proper homothetic vector fields. These symmetries occur in some special cases and will be explained later. It is assumed that the space-times under consideration admit no such symmetries and it is not constant curvature. The non-zero independent components of the Riemann tensor are

$$R^{01}{}_{01} = -\frac{1}{4}e^{-A(t,r)-B(t,r)}\left[\begin{array}{l}e^{A(t,r)}(A_r^2(t,r)+2A_{rr}(t,r))-e^{B(t,r)}(B_t^2(t,r)+2B_{tt}(t,r)-\\ A_t(t,r)B_t(t,r))-e^{A(t,r)}A_r(t,r)B_r(t,r)\end{array}\right] \equiv \alpha_1,$$

$$R^{02}{}_{02} = R^{03}{}_{03} = -\frac{1}{2r}e^{-B(t,r)}A_r(t,r) \equiv \alpha_2, \quad R^{12}{}_{12} = R^{13}{}_{13} = \frac{1}{2r}B_r(t,r)e^{-B(t,r)} \equiv \alpha_3,$$

$$R^{23}{}_{23} = \frac{1}{r^2}(1-e^{-B(t,r)}) \equiv \alpha_4, \quad R^{02}{}_{12} = R^{03}{}_{13} = -\frac{1}{2r}B_t(t,r)e^{-A(t,r)} \equiv \alpha_5, \tag{6}$$

$$R^{12}{}_{02} = R^{13}{}_{03} = -\frac{1}{2r}B_t(t,r)e^{-B(t,r)} \equiv \alpha_6,$$

where $\alpha_1, \alpha_2, \alpha_3, \alpha_4, \alpha_5$ and $\alpha_6$ are real functions of $t$ and $r$ only. One can write the above equation (6) as

$$R^{ab}{}_{cd}{}^1F^{cd} = \alpha_1{}^1F^{ab}, \qquad\qquad R^{ab}{}_{cd}{}^2F^{cd} = \alpha_2{}^2F^{ab}+\alpha_6{}^4F^{ab},$$
$$R^{ab}{}_{cd}{}^3F^{cd} = \alpha_2{}^3F^{ab}+\alpha_6{}^5F^{ab}, \qquad R^{ab}{}_{cd}{}^4F^{cd} = \alpha_3{}^4F^{ab}+\alpha_5{}^2F^{ab},$$
$$R^{ab}{}_{cd}{}^5F^{cd} = \alpha_3{}^5F^{ab}+\alpha_5{}^3F^{ab}, \qquad R^{ab}{}_{cd}{}^6F^{cd} = \alpha_4{}^6F^{ab}.$$

Here, we are interested in the eigenvalues of Riemann tensor to find the proper projective collineation. To make it eigenvalue and eigenvector problem we need to choose $\alpha_5 = 0 \Rightarrow B_t(t,r) = 0 \Rightarrow B = B(r)$ and also we have $\alpha_6 = 0$. It is important to mention here that throughout in this paper we have $B = B(r)$. Here, at $p \in M$ one can choose the tetrad $(t, r, \theta, \phi)$ satisfying $-t^a t_a = r^a r_a = \theta^a \theta_a = \phi^a \phi_a = 1$ (with all other inner products zero) such that the eigenbivectors of the curvature tensor at $p \in M$ are all simple with blades spanned by the vector pairs $(t,\theta)$, $(t,\varphi)$ each with eigenvalue $\alpha_2(p)$ and $(r,\theta)$, $(r,\varphi)$ each with eigenvalue $\alpha_3(p)$. Here the vector fields are chosen as $t_a = e^{\frac{A}{2}}\delta_a^0$, $r_a = e^{\frac{B}{2}}\delta_a^1$, $\theta_a = r\delta_a^2$ and $\phi_a = r\sin\theta\,\delta_a^3$. We are considering the open sub region where $\alpha_2$ and $\alpha_3$ are nowhere equal and $\alpha_2 \neq 0$. The rest will be considered latter. It is important to note that we are using $(t,r,\theta,\phi)$ as both coordinates and vector fields. Thus at $p$, the tensor $P_{ab} = \alpha_2 h_{ab} + 2\eta_{a;b}$ has eigenvectors $t,\theta,\varphi$ with same eigenvalue, say $\beta_1$ and $P_{ab} = \alpha_3 h_{ab} + 2\eta_{a;b}$ has eigenvectors $r,\theta,\varphi$ with same eigenvalue, say $\beta_2$. Hence on $M$ one has after the use of completeness relation $(g_{ab} = -t_a t_b + r_a r_b + \theta_a \theta_b + \phi_a \phi_b)$

$$\alpha_2 h_{ab} + 2\eta_{a;b} = \beta_1 g_{ab} + \beta_3 r_a r_b, \qquad \alpha_3 h_{ab} + 2\eta_{a;b} = \beta_2 g_{ab} + \beta_4 t_a t_b \qquad (7)$$

where $\beta_1, \beta_2, \beta_3$ and $\beta_4$ are real functions on $M$. Since $\alpha_2 \neq \alpha_3$ then it follows from (6) that

$$h_{ab} = Q g_{ab} + D r_a r_b + E t_a t_b, \qquad \eta_{a;b} = F g_{ab} + G r_a r_b + K t_a t_b, \qquad (8)$$

where $D, E, F, G, K$ and $Q$ are functions on some open subregion of $M$. Next one substitutes the first equation of (8) in (2), we get

$$Q_{,c} g_{ab} + D_{,c} r_a r_b + D r_{a;c} r_b + D r_{b;c} r_a + E_{,c} t_a t_b + E t_{a;c} t_b + E t_{b;c} t_a$$
$$= 2 g_{ab} \eta_c + g_{ac} \eta_b + g_{bc} \eta_a. \qquad (9)$$

Contracting the above equation with $\theta^a \phi^b$, and then comparing both sides, we have $\eta_a \theta^a = \eta_a \phi^a = 0$ which implies $\eta = \eta(t,r)$. Now contracting equation (9) with $\theta^a \theta^b$ we get $Q_{,c} = 2\eta_c$ which implies $Q = Q(t,r)$. Once again contracting equation (9) with $t^a t^b$ and $r^a r^b$, we get $E = E(t)$ and $D = D(r)$, respectively. Now consider the first equation of (8) and using (4), we get the following non-zero components of $h_{ab}$

$$h_{00} = [E(t) - Q(t,r)]e^{A(t,r)}, \qquad h_{11} = [Q(t,r) + D(r)]e^{B(r)},$$
$$h_{22} = Q(t,r) r^2, \qquad h_{33} = Q(t,r) r^2 \sin^2 \theta. \qquad (10)$$

Now we are interested in finding the projective vector fields by using the relation

$$L_X g_{ab} = h_{ab}, \qquad \forall\, a,b = 0,1,2,3. \qquad (11)$$

Writing equation (11) explicitly and using (4) and (10) we get

$$A_t(t,r) X^0 + A_r(t,r) X^1 + 2 X^0_{,0} = Q(t,r) - E(t) \qquad (12)$$

$$e^{B(r)} X^1_{,0} - e^{A(t,r)} X^0_{,1} = 0 \tag{13}$$

$$r^2 X^2_{,0} - e^{A(t,r)} X^0_{,2} = 0 \tag{14}$$

$$r^2 \sin^2 \theta X^3_{,0} - e^{A(t,r)} X^0_{,3} = 0 \tag{15}$$

$$B_r(r) X^1 + 2 X^1_{,1} = Q(t,r) + D(r) \tag{16}$$

$$r^2 X^2_{,1} + e^{B(r)} X^1_{,2} = 0 \tag{17}$$

$$r^2 \sin^2 \theta X^3_{,1} + e^{B(r)} X^1_{,3} = 0 \tag{18}$$

$$X^1 + r X^2_{,2} = \frac{r}{2} Q(t,r) \tag{19}$$

$$\sin^2 \theta X^3_{,2} + X^2_{,3} = 0 \tag{20}$$

$$X^1 + r \cot \theta X^2 + r X^3_{,3} = \frac{r}{2} Q(t,r). \tag{21}$$

Considering equations (14) and (15) and differentiating with respect to $\phi$ and $\theta$, respectively and subtracting them we get

$$-\left[\sin^2 \theta X^3_{,0}\right]_{,2} + X^2_{,03} = 0. \tag{22}$$

Differentiating equation (20) with respect to $t$ we get

$$\sin^2 \theta X^3_{,02} + X^2_{,03} = 0. \tag{23}$$

Subtracting equation (22) from equation (23) and integrating we get

$$X^3 = \cos ec\theta \int E^1(t,r,\phi) dt + E^2(r,\theta,\phi),$$

where $E^1(t,r,\phi)$ and $E^2(r,\theta,\phi)$ are functions of integration. Using the above information in equation (15) one has

$$X^0 = e^{-A(t,r)} r^2 \sin \theta \int E^1(t,r,\phi) d\phi + E^3(t,r,\theta),$$

where $E^3(t,r,\theta)$ is a function of integration. Substituting the value of $X^0$ in equations (13) and (14) we get

$$X^1 = r^2 \sin \theta e^{-B(r)} \int \left[\int E^1_r(t,r,\phi) d\phi\right] dt + 2r \sin \theta e^{-B(r)} \int \left[\int E^1(t,r,\phi) d\phi\right] dt$$
$$- r^2 \sin \theta e^{-B(r)} \int \left[A_r \int E^1(t,r,\phi) dt\right] dt + \int e^{A(t,r)-B(r)} E^3_r(t,r,\theta) dt + E^4(r,\theta,\phi),$$

$$X^2 = \cos \theta \int E^1(t,r,\phi) d\phi \, dt + \frac{1}{r^2} \int e^{A(t,r)} E^3_\theta(t,r,\theta) dt + E^5(r,\theta,\phi),$$

where $E^4(r,\theta,\phi)$ and $E^5(r,\theta,\phi)$ are functions of integration. Thus we have

$$X^0 = e^{-A(t,r)} r^2 \sin\theta \int E^1(t,r,\phi) d\phi + E^3(t,r,\theta),$$

$$X^1 = r^2 \sin\theta\, e^{-B(r)} \int \left[\int E^1_r(t,r,\phi) d\phi\right] dt + 2r \sin\theta\, e^{-B(r)} \int \left[\int E^1(t,r,\phi) d\phi\right] dt$$

$$-r^2 \sin\theta\, e^{-B(r)} \int \left[A_r \int E^1(t,r,\phi) d\phi\right] dt + \int e^{A(t,r)-B(r)} E^3_r(t,r,\theta) dt + E^4(r,\theta,\phi), \quad (24)$$

$$X^2 = \cos\theta \int E^1(t,r,\phi) d\phi\, dt + \frac{1}{r^2} \int e^{A(t,r)} E^3_\theta(t,r,\theta)\, dt + E^5(r,\theta,\phi),$$

$$X^3 = \cos ec\,\theta \int E^1(t,r,\phi) dt + E^2(r,\theta,\phi).$$

If one proceeds further after some lengthy calculation one finds that

$$X^0 = \sin\theta \left[K^1(t)\sin\phi + K^2(t)\cos\phi\right] e^{\int \frac{1}{r} e^{B(r)} dr} - \cos\theta\, K^3(t) e^{\int \frac{1}{r} e^{B(r)} dr} + F^4(t,r),$$

$$X^1 = \frac{1}{r} \sin\theta \int \left[K^1(t)\sin\phi - K^2(t)\cos\phi\right] e^{A(t,r)+\int \frac{1}{r} e^{B(r)} dr} dt + \int e^{A(t,r)-B(r)} F^4_r(t,r) dt$$

$$-\frac{1}{r} \cos\theta \int e^{A(t,r)} K^3(t) e^{\int \frac{1}{r} e^{B(r)} dr} dt + E^4(r,\theta,\phi), \quad (25)$$

$$X^2 = \frac{1}{r^2} \cos\theta \int \left[K^1(t)\sin\phi - K^2(t)\cos\phi\right] e^{A(t,r)+\int \frac{1}{r} e^{B(r)} dr} dt +$$

$$\frac{\sin\theta}{r^2} \int K^3(t) e^{A(t,r)+\int \frac{1}{r} e^{B(r)} dr} dt + E^5(r,\theta,\phi),$$

$$X^3 = \frac{1}{r^2} \cos ec\,\theta \int \left[K^1(t)\cos\phi + K^2(t)\sin\phi\right] e^{A(t,r)+\int \frac{1}{r} e^{B(r)} dr} dt + E^2(r,\theta,\phi),$$

where $K^1(t), K^2(t), K^3(t)$ and $F^4(t,r)$ are functions of integration. Differentiating equation (17) with respect to $\phi$ and $t$, and using the above information and equation (25) we have

$$[K^1(t)\cos\phi + K^2(t)\sin\phi][A_r(t,r) + \frac{2}{r} e^{B(r)} - \frac{2}{r}] = 0.$$

The above equation gives the following three possible subcases which are:

**(I)** $[K^1(t)\cos\phi + K^2(t)\sin\phi] = 0$ and $A_r(t,r) + \frac{2}{r} e^{B(r)} - \frac{2}{r} \neq 0$

**(II)** $[K^1(t)\cos\phi + K^2(t)\sin\phi] \neq 0$ and $A_r(t,r) + \frac{2}{r} e^{B(r)} - \frac{2}{r} = 0$

**(III)** $[K^1(t)\cos\phi + K^2(t)\sin\phi] = 0$ and $A_r(t,r) + \frac{2}{r} e^{B(r)} - \frac{2}{r} = 0.$

We discuss each of the above subcases in turn.

**Case (I)**

In this case we have $[K^1(t)\cos\phi + K^2(t)\sin\phi] = 0$ and $A_r(t,r) + \frac{2}{r}e^{B(r)} - \frac{2}{r} \neq 0$.

Equation $[K^1(t)\cos\phi + K^2(t)\sin\phi] = 0 \Rightarrow K^1(t) = K^2(t) = 0$. Using the above information in equation (25) we get

$$\begin{aligned}
X^0 &= F^4(t,r) \\
X^1 &= \int e^{A(t,r)-B(r)} F_r^4(t,r) dt + \sin\theta\, F^1(r,\phi) + F^2(r,\phi) + F^3(r,\theta) \\
X^2 &= \cos\theta\, F^5(r,\phi) + F^6(r,\theta) + F^7(\theta,\phi) \\
X^3 &= E^2(r,\theta,\phi)
\end{aligned} \quad (26)$$

Differentiate (12) with respect to $\theta$ and $\phi$, using (26) we have $A_r(t,r) F_\phi^2(r,\phi) = 0 \Rightarrow A_r(t,r) \neq 0$ and $F_\phi^2(r,\phi) = 0 \Rightarrow F_\phi^2(r,\phi) = 0$ (if $A_r(t,r) = 0 \Rightarrow \alpha_2 = 0$ which gives contradiction to our assumption. Hence $A_r(t,r) \neq 0$.) which implies $F^2(r,\phi) = G^1(r)$, where $G^1(r)$ is a function of integration. If one proceeds further after some straightforward calculations, one finds the solution of equations (12)-(21) (after subtracting Killing vector fields),

$$X^0 = F^4(t,r),\ X^1 = \int e^{A(t,r)-B(r)} F_r^4(t,r) dt + c_1 e^{-\frac{1}{2}B(r)} + G^2(r) \quad (27)$$
$$X^2 = 0,\qquad X^3 = 0$$

provided that

$$A_t(t,r) F^4(t,r) + A_r(t,r)\left[\int e^{A(t,r)-B(r)} F_r^4(t,r) + c_1 e^{-\frac{1}{2}B(r)} + G^2(r)\right] + 2 F_t^4(t,r)$$
$$= Q(t,r) - E(t),$$
$$B_r(r)\left[\int e^{A(t,r)-B(r)} F_r^4(t,r) dt + G^2(r)\right] \quad (28)$$
$$+ 2\int e^{A(t,r)-B(r)}\left[F_{rr}^4(t,r) + F_r^4(t,r)\{A_r(t,r) - B_r(r)\}\right] dt + 2 G_r^2(r) = Q(t,r) + D(r),$$
$$\int e^{A(t,r)-B(r)} F_r^4(t,r) dt + c_1 e^{-\frac{1}{2}B(r)} + G^2(r) = \frac{r}{2} Q(t,r),$$

where $G^2(r)$ is a function of integration and $c_1 \in R$. Suppose $X = (\gamma(t,r), \rho(t,r), 0, 0)$, where $\gamma(t,r) = F^4(t,r)$, $\rho(t,r) = \int e^{A(t,r)-B(t,r)} F_r^4(t,r) dt + c_1 e^{-\frac{1}{2}B(r)} + G^2(r)$ and one form $\eta_a = \lambda_1(t,r) t_a + \lambda_2(t,r) r_a$.

The vector field is then called projective collineation if it satisfies (2). So, using the above information in (2) gives

$$\lambda_1 = 0,\ E = c_3,\ Q = c_4 r^2 + c_3,\ D = c_4 r^2,\ \lambda_2 = r c_4,\ A = \ln\left|p^1(t)(c_4 r^2 + c_3)\right|$$ and

$$B = \ln\left|\frac{c_5}{c_4 r^2 + c_3}\right|,$$ where $P^1(t)$ is a no where zero function of integration and

$c_3, c_4, c_5 \in R (c_4 \neq 0, c_5 \neq 0, c_4 \neq c_5, c_3 \neq c_5)$. *Thus the space-time (4) admits a proper projective collineation, for a special choice of $A$ and $B$ as given above.* Substituting the value of $A$ and $B$ in (4), the space-time can, after a suitable rescaling of $t$ be written in the form

$$ds^2 = -(c_4 r^2 + c_3) dt^2 + \frac{c_5}{c_4 r^2 + c_3} dr^2 + r^2 (d\theta^2 + \sin^2\theta\, d\phi^2) \quad (29)$$

with $\eta_a = (rc_4) r_a$. Proper projective collineation after subtracting Killing vector fields is

$$X = (0, \frac{r}{2}(c_4 r^2 + c_3), 0, 0). \quad (30)$$

One can also find a proper homothetic vector field by setting $Q(\neq 0) \in R$ and $D = E = 0$ in equation (27) and (28) and in fact $A = \ln \alpha r^2$ and $B = $ constant, where $\alpha \in R \setminus \{0\}$. Conversely, if $A = \ln \alpha r^2$ and $B = $ constant, then the space-time (4) admits a proper homothetic vector field which is (after subtracting Killing vector fields) $X = (0, \frac{Q}{2} r, 0, 0)$. Similarly Killing vector fields can be obtained by setting $D = E = Q = 0$ in equations (27) and (28), one can find Killing vector fields which are given in equation (5). It is important to note that the above space-time (29) which admits proper projective collineation is a special class of static spherically symmetric space-time and its Ricci tensor Segre type is $\{(1,1)(11)\}$.

**Case (II)**

In this case we have $[K^1(t)\cos\phi + K^2(t)\sin\phi] \neq 0$ and $A_r(t,r) + \frac{2}{r} e^{B(r)} - \frac{2}{r} = 0$. Substituting the above information back in the same equation and after lengthy calculations one finds that $K^1(t)=0$ and $K^2(t)=0$ which gives a contradiction (since we assume that $[K^1(t)\cos\phi + K^2(t)\sin\phi] \neq 0$). Hence this subcase is not possible.

**Case (III)**

In this case we have $[K^1(t)\cos\phi + K^2(t)\sin\phi]=0$ and $A_r(t,r)+\frac{2}{r}e^{B(r)}-\frac{2}{r}=0$.

Equation $[K^1(t)\cos\phi + K^2(t)\sin\phi]=0 \Rightarrow K^1(t)=0$ and $K^2(t)=0$. If one proceeds further after some lengthy calculations, one finds the solution of equations (12)–(21) (after subtracting Killing vector fields),

$$X^0 = F^4(t,r),\ X^1 = \int e^{A(t,r)-B(r)} F_r^4(t,r)dt + G^2(r)$$
$$X^2 = 0, \qquad X^3 = 0 \tag{31}$$

provided that

$$A_t(t,r)F^4(t,r) + A_r(t,r)\left[\int e^{A(t,r)-B(r)} F_r^4(t,r) + G^2(r)\right] + 2F_t^4(t,r)$$
$$= Q(t,r) - E(t),$$
$$B_r(r)\left[\int e^{A(t,r)-B(r)} F_r^4(t,r)dt + G^2(r)\right] \tag{32}$$
$$+ 2\int e^{A(t,r)-B(r)} \left[F_{rr}^4(t,r) + F_r^4(t,r)\{A_r(t,r)-B_r(r)\}\right]dt + 2G_r^2(r) = Q(t,r) + D(r),$$
$$\int e^{A(t,r)-B(r)} F_r^4(t,r)dt + G^2(r) = \frac{r}{2}Q(t,r),$$

where $G^2(r)$ is a function of integration. Suppose $X=(\gamma(t,r),\rho(t,r),0,0)$, where $\gamma(t,r)=F^4(t,r),\ \rho(t,r)= \int e^{A(t,r)-B(t,r)} F_r^4(t,r)dt + G^2(r)$ and one form $\eta_a = \lambda_1(t,r)t_a + \lambda_2(t,r)r_a$. The vector field is then called projective collineation if it satisfies (2). Using the above information in (2) gives

$$\lambda_1 = 0,\ E=c_6,\ Q=c_7 r^2 + c_6,\ D=c_7 r^2,\ \lambda_2 = rc_7,\ A = \ln\left|p^2(t)(c_7 r^2 + c_6)\right|,$$

$$B = \ln\left|\frac{c_6}{c_7 r^2 + c_6}\right| \text{ and } \eta_a = (rc_7)r_a. \tag{33}$$

where $P^2(t)$ is a no where zero function of integration and $c_6, c_7 \in R(c_6 \neq 0, c_7 \neq 0, c_6 \neq c_7)$. *Thus the space-time (4) admits a proper projective collineation, for a special choice of A and B as given above in* *equation (33).* Substituting the value of $A$ and $B$ in (4), the space-time can, after a suitable rescaling of $t$ be written in the form

$$ds^2 = -(c_7 r^2 + c_6)dt^2 + \frac{c_6}{c_7 r^2 + c_6}dr^2 + r^2(d\theta^2 + \sin^2\theta d\phi^2) \quad (34)$$

with $\eta_a = (rc_7)r_a$. Proper projective collineation after subtracting Killing vector fields is

$$X = (0, \frac{r}{2}(c_7 r^2 + c_6), 0, 0). \quad (35)$$

The above space-time (34) which admits proper projective collineation turns out to be special class of static spherically symmetric space-time and its Ricci tensor Segre type is $\{(1,1)(11)\}$.

Now consider the case when $\alpha_2 = 0$ in (6) then one can see the rank of the $6 \times 6$ Riemann tensor is 3 or $R^a{}_{bcd}t^d = 0$, where $t^a$ is a timelike vector field and unique solution of $R^a{}_{bcd}t^d = 0$. Here, it is important to mention here that $B = B(r)$ and $\alpha_2 \neq \alpha_3$. The condition $\alpha_2 = 0 \Rightarrow A_r(t,r) = 0$ which implies $A = A(t)$. The line element (4) can, after a rescaling of $t$, be written in this form

$$ds^2 = -dt^2 + e^{B(r)}dr^2 + r^2(d\theta^2 + \sin^2\theta d\phi^2). \quad (36)$$

The above space-time is 1+3 decomposable. It follows from [2] that space-time (36) does not admit proper projective collineation. The projective collineation admitted by (36) is a proper affine vector field which is $tt^a$.

Consider when $\alpha_2 = \alpha_3$ (and excluding the special case when $A = $ constant and $B = $ constant $\neq 0$) in (4). It follows from [4,8] that projective collineation admitted by (4) are Killing vector fields which are given in (5).

Now consider the special case when $A = $ constant and $B = $ constant $\neq 0$. The rank of the $6 \times 6$ Riemann tensor is 1 and there exists two independent solutions, which are $R^a{}_{bcd} t^d = 0$ and $R^a{}_{bcd} r^d = 0$, but only one independent covariantly constant vector field $t_a = t_{,a}$ satisfying $t_{a;b} = 0$. Substituting the above information in (4) and after a rescaling of $t$, the line element takes the form

$$ds^2 = -dt^2 + k\, dr^2 + r^2(d\theta^2 + \sin^2\theta d\varphi^2) \quad (37)$$

where $k(= e^B) \in R(k \neq 0 \, or \, 1)$. The Ricci tensor Segre type of the above space time (37) is $\{(1,1)(11)\}$. The space time is clearly 1+3 decomposable, but the rank of the $6 \times 6$ Riemann tensor is 1. Here one can clearly see that the space-time (37) which admits proper special projective collineation is a special class of static spherically symmetric space-time. We know from [2,5] that the above space time (37) admits a unique (up to an affine) proper special projective collineation, proper affine vector fields and proper homothetic vector fields which are:

$$\begin{aligned} U &= (t^2, tr, 0, 0), & V &= (t, 0, 0, 0) \\ Z &= (0, r, 0, 0) & T &= (t, r, 0, 0), \end{aligned} \quad (38)$$

where U is a proper special projective collineation, V and Z are proper affine vector fields and T is a proper homothetic vector field. The dimension of special projective collineation is 7.

## ACKNOWLEDGMENT

The author (GS) would like to thank Prof. G S Hall (University of Aberdeen, UK) for many helpful and interesting discussions.